# Spectral and Small-Signal Electroluminescence Analysis of Carrier Dynamics in Dual-Color InGaN/GaN Light-Emitting Diodes

Xuefeng Li[1*], Rob Armitage[2], and Daniel Feezell[1]

[1]*Center for High Technology Materials (CHTM), University of New Mexico, Albuquerque, NM 87106, USA*
[2]*Lumileds LLC, San Jose, CA 95131, USA*

**Abstract:** We study carrier transport, distribution, and recombination in dual-color *c*-plane InGaN/GaN LEDs using spectral analysis and small-signal electroluminescence (SSEL). The emissions from green and blue quantum wells (QWs) were experimentally separated and analyzed. Spectral analysis and SSEL independently demonstrate that emission from the green QW is dominant at low current densities due to its narrower bandgap, while emission from the blue QW is more significant at higher current densities due to its reduced quantum confined stark effect (QCSE) and larger wavefunction overlap. In addition, we demonstrate that the carrier recombination in the QWs is non-uniform, with carrier transport dramatically affecting the carrier distribution between QWs and the recombination in a specific QW. Finally, we also show that the effective active region in these V-pit-engineered InGaN/GaN LEDs is roughly 2 to 3 QWs on the p-GaN side, with limited interwell carrier transport and recombination in additional QWs.

[*]**Electronic mail:** xuefengli@ucsb.edu

1. **Introduction**

Continually improving the quantum efficiency of long-wavelength InGaN/GaN light-emitting diodes (LEDs) is important for solid-state lighting and full-color RGB (red-green-blue) displays [1]. Due to strong hole confinement and slow carrier transport in multiple quantum wells (MQWs) [2-3], interwell carrier transport is limited, with carriers mostly confined to the QW next to the p-GaN [4-7]. Enhancing carrier transport and achieving uniform emission across MQWs will help lower the carrier density ($n$) in the QWs and reduce Auger-Meitner recombination, the dominant nonradiative recombination process at high $n$ in commercial-grade green InGaN/GaN LEDs [8]. In single-color MQW InGaN/GaN LEDs, it is challenging to experimentally identify the contributions of specific QWs to photon emission and to study the carrier transport dynamics. Therefore, understanding the carrier distribution and recombination dynamics within MQWs is crucial for optimizing the active region and improving the quantum efficiency in LEDs. Alternatively, dual-color LEDs [9] enable the study of interwell carrier dynamics by emission spectrum decomposition, which allows for targeted analysis of the carrier distribution and transport for specific QWs, making dual-color LEDs excellent testbeds for carrier dynamics and quantum efficiency studies.

To address the issues related to strong hole confinement and limited carrier transport within MQWs, a three-dimensional defect-controlled approach using V-pit engineering [10-11] has been employed. V-pits are open hexagonal inverted pyramid-shaped defects [12-13] that introduce additional vertical and lateral carrier transport paths. The additional paths are beneficial for interwell carrier transport due to the thin barrier layer and low indium content on the semipolar $\{10\bar{1}1\}$ plane [14-18]. By enhancing carrier transport, V-pit engineering increases the effective active region volume, reduces carrier escape, helps maintain a low $n$ in the active region, and enhances the quantum efficiency [19-24]. Therefore, a study of the carrier dynamics in dual-color InGaN/GaN LEDs is informative for quantum efficiency improvement in long-wavelength InGaN/GaN LEDs, such as green, amber, and red.

Recent studies have demonstrated that QW emission is concentrated around V-pits in green/red dual-color LEDs using near-field EL [25], and that there are different hole injection paths in structures with and without such structures [26]. Here, we use EL and SSEL to examine the carrier dynamics of emitting QWs at different current densities, the carrier distribution among the QWs, and the effective QW number or active region volume in dual-color LEDs by comparing similar epitaxial structures but with different QW designs. Due to the complex carrier dynamics in the dual-color QW region, we are unable to extract definitive carrier transport and recombination lifetimes. Nonetheless, by comparing

SSEL (and EL) characteristics among the studied dual-color LEDs, we still obtain insights into carrier transport, distribution, and recombination behaviors. For example, EL intensity is informative for carrier distribution and recombination. In general, a higher EL intensity indicates a higher carrier recombination rate and a higher carrier density. SSEL also provides direct evidence for effective recombination dynamics. A broader modulation bandwidth generally corresponds to a shorter differential recombination lifetime. Additionally, the relative change of the emission intensity between different colors as a function of current density provides indirect information about carrier redistribution among QWs. Overall, these comparisons provide valuable insight into the carrier dynamics in dual-color LEDs, even without determining the exact carrier lifetimes.

## 2. Experiment details

In this work, we study four *c*-plane dual-color InGaN/GaN LEDs grown with state-of-the-art growth conditions using metal-organic chemical vapor deposition (MOCVD). The four dual-color LEDs share the same epitaxial structure, except for the differences in the QW region. There are either two or four QWs in each wafer, and they were designed to emit blue or green light. The corresponding indium compositions for blue and green QWs are 16% and 19% based on energy-dispersive x-ray spectroscopy (EDX) measurements. The epitaxial structures of the LEDs are given and referred to as follows: n-GaN/blue QW/green QW/p-GaN ("n/b/g/p"), n-GaN/green QW/blue QW/p-GaN ("n/g/b/p"), n-GaN/2X blue QW/2X green QW/p-GaN ("n/2b/2g/p"), and n-GaN/2X green QW/2X blue QW/p-GaN ("n/2g/2b/p"), as illustrated in Figure 1. The QWs all have a thickness of 3 nm, and the GaN barriers are each 18 nm thick.

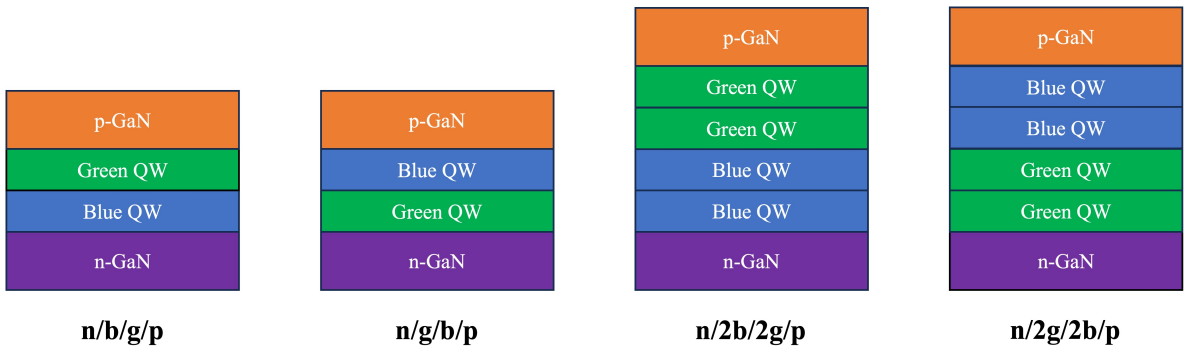

Fig. 1. Epitaxial structure for dual-color InGaN/GaN LEDs.

## 3. Results and discussions

Figure 2 presents the normalized electroluminescence (EL) intensities for the dual-color LED series. There is a noticeable blue shift in the EL for both blue and green emission at high current density ($J$) [27-29]. In the n/b/g/p LED (Figure 2(a)), the green QW is the primary photon-emitting QW, especially at low current densities. However, blue emission is also observed when $J$ exceeds 10 A/cm$^2$, suggesting that carriers reach to and recombine within the QW on the n-GaN side. This emission behavior from the n-GaN side QW is different from single-color blue, cyan, or green InGaN/GaN LEDs with identical QW thickness, indium composition, and barrier thickness but without intentional V pits growth and no enhancement of carrier transport, where photon emission is typically concentrated in the QW next to p-GaN [30]. Even without intentional growth modification, V-pits may still form because threading dislocations (TDs) terminate at the growth surface as a result of the large lattice mismatch between GaN and the substrate and the associated strain during epitaxy [18, 31-32]. For simplicity, LEDs in which V-pits form only unintentionally are referred to as "without V-pits" in this work. In the blue QW, the carrier recombination rate is faster than in the green QW at a given $n$, due to a weaker quantum-confined Stark effect (QCSE) and greater wavefunction overlap [33-37]. A similar behavior is also observed in the n/2b/2g/p (Figure 2(c)) LED, where photon emission from the blue QWs appears when $J$ exceeds 20 A/cm$^2$, though the emission is less strong than in the n/b/g/p LED. In the n/2b/2g/p LED, the separation between the green and blue QWs is larger, which may contribute to the reduced blue emission intensity.

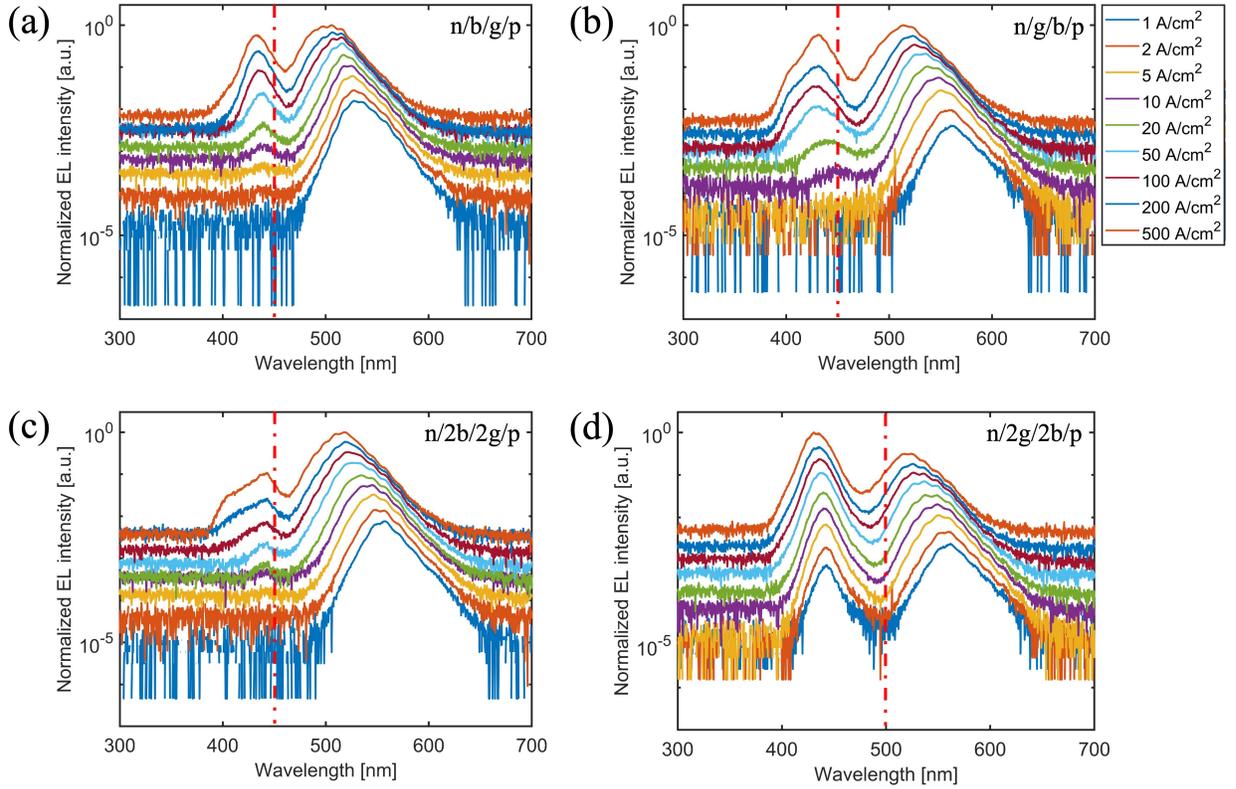

Fig. 2. Normalized EL intensity for (a) n/b/g/p, (b) n/g/b/p, (c) n/2b/2g/p, (d) n/2g/2b/p wafers at different $J$.

In the n/g/b/p LED (Figure 2(b)), the blue QW next to the p-GaN is expected to be the primary QW for carrier recombination for LEDs without V pits. However, there is no blue emission detected below 10 A/cm$^2$, while strong green emission is observed at low $J$. The reabsorption process of blue photons in green QWs has only a limited effect on the emission spectrum, as was previously reported in photoluminescence studies [38]. In later sections, we will also provide further evidence of a weak reabsorption process in the InGaN/GaN LEDs studied here. Therefore, the narrower bandgap of the green QW is associated with a lower turn-on voltage, which correlates with strong green emission at low $J$. Apparently, the effect of narrower bandgap outweighs the impact of stronger QCSE, reduced wavefunction overlap, and significant hole confinement. Similarly, strong green emission is also observed at low $J$ in the n/2g/2b/p LED (Figure 2(d)). However, there is also strong blue emission detected at low $J$ in the n/2g/2b/p LED, indicating a complex interplay among the effect of QCSE, wavefunction overlap, hole confinement, and the narrower bandgap of the green QWs.

Next, we study the number of photons emitted by different color QWs. As shown in Figure 3(a), the normalized EL spectra of blue and green QWs overlap. Here, we use the normalized EL intensity of the n/b/g/p LED at 500 A/cm² to illustrate the separation between blue and green emissions. To minimize the effects of Fabry–Pérot resonance [39] on the analysis, a Fast Fourier Transform (FFT) [40-41] was applied to the raw normalized EL data with a cutoff frequency of 60 kHz. We then model the EL spectrum of blue and green QWs using two Gaussian functions, represented by blue and green open dots, respectively, and fit them to the FFT-filtered EL spectrum. The combined EL spectrum of both QWs, shown by black filled dots, aligns well with the FFT-filtered EL data. The wavelength at the dip between the blue and green emission spectra (around 462 nm) effectively separates the emissions from blue and green QWs. In the analysis, we define the cut-off wavelength ($\lambda_{cut-off}$) as the normalized dip in the number of photons between blue and green emission spectra.

The EL energy ($E$) at a given wavelength is:

$$E = n(\lambda)h\nu = n(\lambda)h\frac{c}{\lambda}$$

Here, $n(\lambda)$ is photon number at a given wavelength $\lambda$, $h$ is Planck constant, $\nu$ is photon frequency, $c$ is the speed of light, and $\lambda$ is photon wavelength.

Then, $n(\lambda)$ can be expressed as:

$$n(\lambda) = \frac{E\lambda}{hc}.$$

Since the EL intensity measured in the EL spectrum is proportional to the EL energy, the total emitted photon number ($N$) over a wavelength range (from $\lambda_1$ to $\lambda_2$) for the LED is:

$$N \propto \int_{\lambda_1}^{\lambda_2} n(\lambda)d\lambda.$$

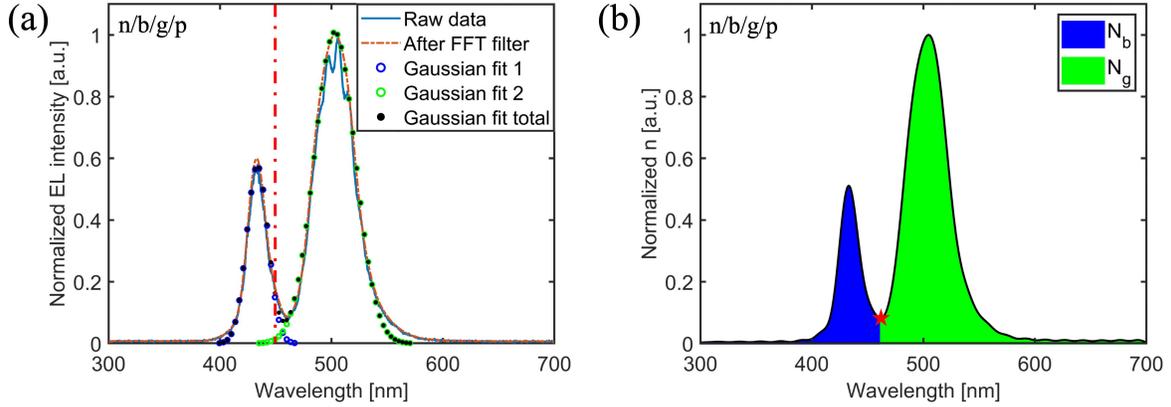

Fig. 3. (a) Normalized EL intensity and (b) $N_b$ and $N_g$, for n/b/g/p LED at 500 A/cm$^2$. The vertical dashed line in Figure 3(a) illustrates the wavelength for bandpass filters.

Thus, $N_b \propto \int_{300\,nm}^{\lambda_{cut-off}} n(\lambda)d\lambda$ and $N_g \propto \int_{\lambda_{cut-off}}^{700\,nm} n(\lambda)d\lambda$, where $N_b$ and $N_g$ are the shadowed areas in Figure 3(b). The red star represents the $\lambda_{cut-off}$ at 462 nm. We then calculate the ratio of photons generated in the blue and green QWs $N_b/N_g$, as shown in Figure 4. A higher percentage of blue photon emission is generally observed at a higher $J$. In the n/b/g/p and n/g/b/p wafers, the increase in blue emission is approximately an order of magnitude more significant than the increase in green emission from 20 A/cm$^2$ to 500 A/cm$^2$. A similar increase in blue emission is observed in the n/2b/2g/p and n/2g/2b/p wafers. This is attributed to the benefit of weaker QCSE, greater wavefunction overlap, and faster carrier recombination rates in the blue QWs. Additionally, there are more blue than green photons from the n/2g/2b/p LED above 100 A/cm$^2$ since $N_b/N_g$ is more than one, suggesting that carrier redistribution has a more significant effect than the bandgap difference at high $J$.

The behavior of $N_b/N_g$ shown in Figure 4 provides additional evidence for a weak reabsorption process in these InGaN/GaN LEDs. If reabsorption were a dominant process, the percentage of blue emission would decrease at a higher $J$, especially for n/b/g/p and n/2b/2g/p wafers, as the blue photons propagate through the green QWs before being detected. This conclusion also applies to n/g/b/p and n/2g/2b/p wafers, where roughly 50% of the blue photon emission (downward emission) passes through the green QWs twice with the topside light collection scheme used here.

Here, we discuss possible reasons for the strong green emission observed in the n/g/b/p and n/2g/2b/p structures, even though their green QWs are located farther from the p-GaN side. At low current density, the applied forward bias is relatively small. There exists a bias range where the quasi-Fermi level separation is sufficient to exceed the green QW bandgap but still below the effective bandgap of the blue QWs. In this regime, radiative recombination occurs predominantly in the green QWs. In addition, the higher indium content in the green QWs leads to deeper potential wells and stronger carrier localization, which enhances carrier capture and radiative efficiency compared with the blue QWs. The presence of V-pits may also facilitate hole injection into the green wells (even though they are farther from the p-GaN side) via semipolar planes. These combined effects explain why the green QWs can emit more strongly even though they are located farther from the p-GaN side in the n/g/b/p and n/2g/2b/p structures.

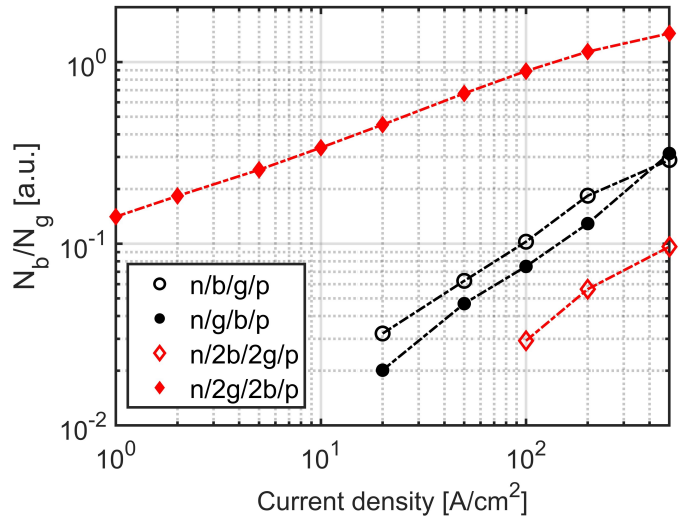

Fig. 4. $N_b/N_g$ for the dual-color LED series.

It is crucial to understand the carrier dynamics across different QWs. Thus, we also study the carrier dynamics in the dual-color LEDs using a small-signal electroluminescence (SSEL) technique [42-43]. The details of the SSEL setup and measurement procedures can be found in Ref. 30. As demonstrated previously, the minimum in the spectrum around 462 nm effectively separates the emissions from the blue and green QWs. To achieve this separation in the SSEL measurements and collect light from only the blue or only the green QWs, 450 nm shortpass and longpass filters (and 500 nm shortpass and longpass filters for n/2g/2b/p wafer to better separate blue and green emissions) were incorporated into a fiber-to-fiber U-Bench coupling setup. Here, shortpass and longpass filters selectively transmit photons below or above a specific wavelength, respectively, while blocking photons outside their designated range. The vertical dashed red lines in Figures 2 and 3(a) indicate the filter cutoff wavelengths. Due to the low coupling efficiency of LED light with the fiber-to-fiber U-Bench setup, we used an avalanche photodiode (APD430A2, Thorlabs) for the measurements. The normalized modulation response (S21) data presented in Figure 5 shows the modulation response of the n/b/g/p LED measured with a 450 nm longpass filter. As shown in Figure 5, the S21 roll off for the n/b/g/p LED with longpass filter is -17 dB/decade, meaning a 17 dB decrease in the signal when the modulation frequency increases by one order of magnitude. This roll off is consistent for all four of the dual-color LEDs measured here with shortpass and longpass filters. This result differs from the typical -20 dB/decade roll off observed in quasi-SQW LEDs. The slower S21 roll off is indicative of an uneven carrier distribution in any specific QW due to strong interwell carrier transport [22]. More specifically, -20 dB/decade roll off corresponds to a uniform QW region rather than a necessarily isolated QW. Unlike in quasi-SQW LEDs, an isolated QW that involves interwell carrier transport cannot be treated as a uniform carrier recombination region based on the -17 dB/decade roll off observed here. S21 roll-off is a direct signature of carrier distribution and lifetime behavior. Carriers may be evenly distributed and thus can be represented by a single lifetime, or they may be unevenly distributed, in which case they must be modeled and analyzed using multiple carrier lifetimes.

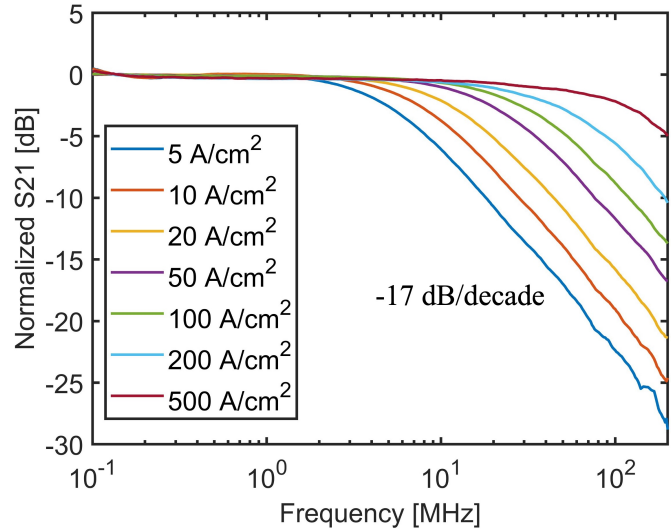

Fig. 5. Modulation response of the n/b/g/p LED measured with APD setup.

Due to the complexity in modeling the carrier dynamics in dual-color LEDs, it is challenging to obtain accurate differential carrier lifetime parameters. As demonstrated in the multiple-carrier-lifetime-model (MCLM) in Ref. 22, -3 dB bandwidth in a QW region is directly associated with the effective differential carrier lifetime in that QW region. The modulation response of an active region with uneven carrier distribution is composed of multiple modulation responses of sub-regions with even carrier distribution. Therefore, -3 dB bandwidth is still a reasonable indicator of effective differential carrier lifetime in dual-color wafers. A larger -3 dB bandwidth implies faster carrier recombination and shorter differential recombination lifetime in a particular region.

Figure 6 presents the -3 dB bandwidth measurements for the dual-color wafers. In n/b/g/p, n/g/b/p, and n/2b/2g/p wafers, the green emission exhibits a wider -3 dB bandwidth at a given $J$, suggesting a higher carrier density in green QWs compared to blue QWs. However, in the n/2g/2b/p wafer, blue emission has a larger -3 dB bandwidth at a given $J$, showing that the carrier density in the blue QW is comparable to or even higher than that in the green QW.

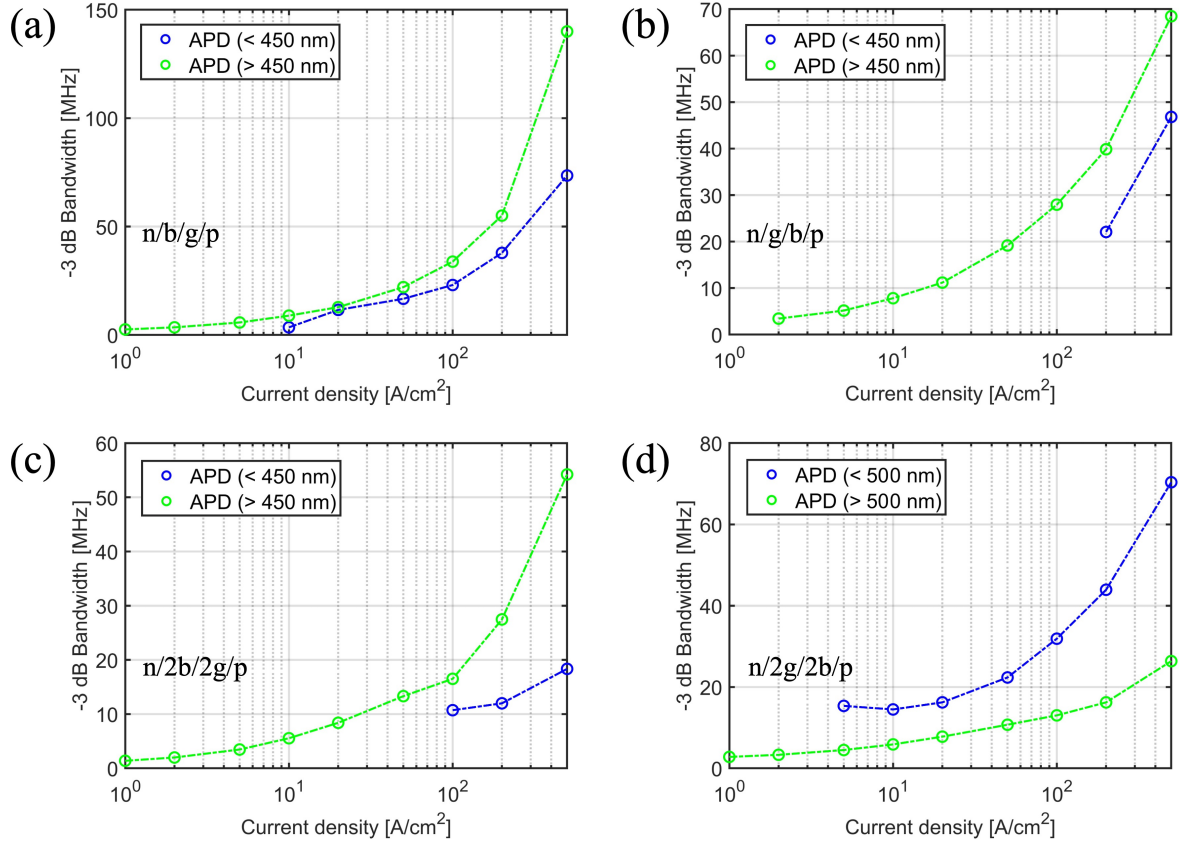

Fig. 6. (a) -3 dB bandwidth of the dual-color LED series.

Comparing the n/2b/2g/p wafer with the n/b/g/p wafer, we observe a decline in the -3 dB bandwidth for both blue and green QWs due to the enlarged active region for both colors. The reduced -3 dB bandwidth and low intensity for the blue emission in the n/2b/2g/p wafer suggest that, despite improved carrier transport compared to LEDs with no V pits, only a limited number of carriers can transport to the far-end (n-GaN side) of the active region. However, the decrease in -3 dB bandwidth for green emission in the n/2b/2g/p wafer indicates strong carrier transport between the green QWs.

Compared with the n/g/b/p wafer, the -3 dB bandwidth increases in blue QWs but decreases in green QWs in the n/2g/2b/p wafer. The advantage of narrower bandgap facilitates carrier distribution in the green QWs, resulting in a larger -3 dB bandwidth for green emission from the QW next to the n-GaN, compared to blue emission from the QW next to the p-GaN in the n/g/b/p wafer. However, the -3 dB bandwidth of the blue emission is much higher than that of the green emission in the n/2g/2b/p wafer, indicating that the effect of improved carrier transport achieved through

V-pit engineering is not particularly effective at injecting carriers into the 3rd QW from the p-GaN side in the LEDs studied here. In the two comparisons, the highly efficient active region volume approximately consists of 2-3 QWs in these LEDs. Carrier transport is weakened by the thick barriers used in our QW designs, and a larger effective QW number and active region volume could be achieved with optimized structures.

4. Conclusion

In summary, we studied the carrier dynamics in a series of dual-color LEDs using EL spectrum and SSEL measurements. We demonstrated that higher-indium-composition QWs tend to have stronger carrier recombination in dual-color LEDs due to their narrower bandgap at low $J$. However, due to the weak QCSE and large wavefunction overlap, low-indium-composition QWs are more favorable for photon emission at high $J$. We showed that carrier recombination characteristics are non-uniform within a given QW under strong interwell carrier transport, i.e., a specific QW is not necessarily a region with uniform carrier recombination rates and lifetimes. We also demonstrated that although V-pit engineering is an effective approach to improve carrier transport, the effective active region contributing strongly to recombination appears to involve approximately 2–3 QWs in the LEDs studied.

**Funding.** Department of Energy (under Award No. DE-EE0009163).

**Disclosures.** The authors declare no conflicts of interest.

**Data availability.** Data underlying the results presented in this paper are not publicly available at this time but may be obtained from the authors upon reasonable request.